\documentclass[12pt,preprint]{aastex}

\shorttitle{Wolf-Rayet Runaway WR 124}
\shortauthors{Marchenko et al.}

\begin{document}

\title{Population I Wolf-Rayet Runaway Stars: the Case of WR124 and its 
Expanding Nebula M1--67}

\author{S.V. Marchenko\altaffilmark{1}, A.F.J. Moffat\altaffilmark{2}, 
P.A. Crowther\altaffilmark{3}}

\altaffiltext{1}{Science Systems and Applications, Inc., 10210 Greenbelt 
Road, Suite 600, Lanham, MD 20706, USA; sergey.marchenko@ssaihq.com}

\altaffiltext{2}{D\'epartement de Physique and Observatoire du Mont 
M\'egantic, Universit\'e de Montr\'eal, CP 6128, Succursale Centre-Ville, 
Montr\'eal, QC H3C 3J7, Canada; moffat@astro.umontreal.ca}

\altaffiltext{3} {University of Sheffield, Department of Physics \& 
Astronomy, Hicks Building, Hounsfield Rd, Sheffield, S3 7RH, UK; 
Paul.Crowther@sheffield.ac.uk}

\begin{abstract} In 1997 and 2008 we used the WFPC2 camera on board of the Hubble
Space Telescope to 
obtain two sets of narrow-band H$\alpha$ images of the runaway Wolf-Rayet 
(WR) star WR 124 surrounded by its nebula M1--67. This two-epoch imaging 
provides an expansion parallax and thus a practically assumption-free 
geometric distance to the nebula, $d=3.35\pm 0.67$ kpc. Combined with the 
global velocity distribution in the ejected nebula, this confirms the 
extreme runaway status of WR 124.  WR stars embedded within such ejection nebulae, at the point of core-collapse would
produce different  supernova characteristics from  those expected for 
stars surrounded by wind-filled cavities.  In galaxies with extremely low ambient metallicity, $Z\leq 10^{-3}
Z_\odot$,  $\gamma$-ray bursts originating from fast-moving runaway Wolf-Rayet stars may produce afterglows
which appear to be coming from regions with a relatively homogeneous circumburst medium.

\end{abstract}


\section{INTRODUCTION}

Among massive stars, those with the largest, continuous mass-loss rates 
are the Population I Wolf-Rayet (WR) stars.  The impact of these fast, 
dense, clumpy and enriched winds on the ISM and stellar evolution itself 
cannot be overstated (e.g., Leitherer et al. 1992). The combination of 
high luminosity and sensitivity to initial relative heavy-element content 
makes WR stars excellent tracers of the ambient metallicity gradients in 
star-forming galaxies. The WR stage is believed to 
precede supernova and hypernova explosions of type Ib/c (Crowther 2007), 
with a  potential link  to the long Gamma-Ray Bursts (GRB; e.g., Mirabal 
et al. 2003).

However, practically all distances to Galactic WR stars are notoriously 
uncertain. Any traditional distance estimates based on radial velocities 
of WR stars with subsequent application of the Galactic rotation law are 
always met with a good degree of skepticism: the optically thick, rapidly 
outmoving envelopes of WR stars distort any information about systemic 
motions of the stellar cores. Hence, modern distances are mainly based on 
spectroscopic parallaxes which, for an individual star, may appear to have 
relatively small errors, $\sim$30\% in the distance and perhaps even lower 
for WR stars in clusters/associations and in some binaries. However, these 
spectroscopic parallaxes are fraught with difficult-to-estimate systematic 
errors, making it essential to find other more direct techniques for 
distance determinations. With regard to the preferable, more robust 
astrometric approach: so far, only one parallax estimate has been made for 
the closest WR star, $\gamma$ Vel (van der Hucht et al. 1997; revised by 
van Leeuwen 2007; the latter being supported by the interferometric 
measurements of North et al. 2007). All remaining WR stars have, so far, 
only lower limits to their parallax-based distances.

The situation is more appalling for the Galactic WN8 stars, i.e., for the 
spectral subclass of our target, WR 124 (209 BAC). Due to the tendency of 
WN8 stars to avoid clusters, associations and binary systems (see 
Marchenko et al. 1998 and references therein), one has to rely, almost 
exclusively, on spectroscopic parallaxes, thus introducing a large 
systematic uncertainty stemming from the necessarily simplified models of 
WR winds. One possible scenario to explain the apparently enigmatic 
characteristics of WN8 stars is that they are runaways, either ejected via 
slingshot-type dynamical interaction from the cores of forming dense star 
clusters (e.g., Evans et al. 2010), or accelerated by a SN explosion of 
the initial primary component in a binary system (De Donder et al. 1997). 
WR124 is of particular interest in this respect, given its large 
heliocentric radial-velocity (RV) component, $v=137$ km s$^{-1}$, based on 
its 
associated nebular H$\alpha$ radial velocity (Sirianni et al 1998).  As 
such, WR 124 is regarded to be among the fastest moving massive stars in 
the Galaxy (Moffat et al. 1982; see also the discussion in Moffat et al. 
1998). The conclusion about the runaway status of WR 124 also heavily 
relies on the rather uncertain estimates of distance provided by (a) the 
fine structure components in the optical interstellar medium (ISM) lines 
seen towards WR 
124 ($d\sim$ 4--5\,kpc: Crawford \& Barlow 1991), and (b) the typical 
luminosities of similar-class WR stars in the Large Magellanic Cloud, 
leading to $d=3.36$ \,kpc. Both estimates result in a high elevation of 
WR 124 above the Galactic plane, $z \geq$ 200 pc, as compared to the 
average $|z|=49$ pc for Galactic WR stars (van der Hucht 2001).

Firmly establishing the runaway nature of WR 124 may have an immediate 
impact on the problem of the environments of GRBs: in spite of the 
expectation  of a $\sim r^{-2}$ density fall-off for a terminal-speed 
stellar wind, at  least 1/5 of GRB afterglows indicate a constant-density 
medium around the  exploding star (van Marle et al. 2006; Starling et al. 2008). An extremely 
rapid motion ($\sim 100$ km s$^{-1}$) of a WR progenitor should lead, on 
an evolutionary timescale, to a displacement of the WR star from its place 
of birth (usually, a star-forming region), as well as from the center of a 
bubble blown by the predecessor's wind, i.e. deep into the surrounding 
and, presumably, more uniform ISM. This is in line with the 
findings of Hammer et al. (2006) that some long GRBs  may occur several 
hundred parsecs  outside  of the  rich and compact clusters or 
star-forming regions whence  they presumably originated, thus defying the 
{\it general} tendency of  GRBs to concentrate on  the brightest regions 
of their host galaxies  (Svensson et al. 2010). Considering the possible 
link of WR stars and, in particular, WR runaways 
to GRBs, one should also be aware that WN8 stars belong to the late-type 
WN (WNL) subgroup, which comprises the most massive and luminous 
Population I  WR  stars. Moreover, while there is considerable variation 
in hydrogen content  from one WN8 star to another, some WN8 stars are 
practically devoid of  hydrogen (see Marchenko et al. 1998 and references 
therein).  Note that WR 124 is further sub-classified as WN8h and so 
contains a non-negligible amount of atmospheric hydrogen (Crowther et al.
1999, C99).

There is an additional aspect which makes WR 124 an attractive target for 
a detailed study. More than 1/4 of all WR stars are currently surrounded 
by detectable nebular shells, seen projected in the form of ring-like 
structures, i.e. ``ring nebulae'' (RNe; Chu et al. 1983, Marston et al. 
1994).  All WR stars have likely possessed a RN at some stage in their 
evolution (Dopita et al. 1994). RNe are believed to be formed via the 
interaction between a slow wind of the predecessor and a fast wind 
associated with the presently observed Population I WR star (e.g. Marston 
1995). Our target belongs to the RN group. The surrounding nebula, M 
1--67, is a relatively young ejection-type, low-excitation nebula (Chu et 
al. 1983; Smith 1995), probably in the earliest [so far] observed phase of 
wind interaction around any Population I WR star.  This is compatible with 
the relatively  high H abundance of the central star. The narrow-band net H$\alpha$ 
image of M1--67 acquired with HST/WFPC2 in 1997 (Grosdidier et al., 1998) 
showed a wealth of complex details, some of which have never been seen 
before in such a nebula. In an attempt to derive a relatively 
assumption-free distance to WR 124/M1--67 based on nebular expansion, we 
obtained a second-epoch image of the nebula in 2008.

\section{OBSERVATIONS AND DATA REDUCTION}

On March 16, 1997, the HST/WFPC2 camera was used to secure a net-H$\alpha$ 
(F656N) image of M1--67 (Grosdidier et al., 1998, 2001). The combination 
of four 2500 sec exposures 
provided a deep, high signal-to-noise Epoch-1 (E1) image. The Epoch-2 (E2) 
HST/WFPC2 image was constructed from 8 500-600 sec-long exposures in 
H$\alpha$ (F656N) obtained on June 21, 2008 in the 4-point dither pattern 
with the cosmic-ray split option. The chosen total exposure time, though 
substantially shorter than in the E1 imaging, nevertheless assured 
adequately high S/N ratios for all parts of the nebula, in addition to 
reducing the impact of the saturated regions surrounding the central star.

All the 1997 and 2008 images were processed using standard {\sc 
iraf}\footnote{{\sc iraf} is distributed by the National Optical Astronomy 
Observatory, which is operated by the Association of Universities 
for Research in Astronomy, Inc. under cooperative agreement with 
the National Science Foundation.} utilities, in particular the {\it 
stsdas/analysis/dither} subroutines. The relative orientation of the E1 
and E2 images, with the latter rotated by $\sim$90 degrees due to the 
technical limitations posed by the 2-gyro guidance mode, prevented a 
straightforward, camera-to-camera comparison of the images. For 
inter-comparison of E1 and E2 we chose 5 regions well distributed 
azimuthally around the central star, where the presence of strong, sharp 
features allowed the best determination of the nebular expansion rate. 
Three of them are sampled by the W3 camera: in one case the central star 
and its immediate surroundings are imaged by W3 in E1 and E2 (Fig. 1); in 
the second case the overlapping regions are imaged by W2 in E1 and W3 in 
E2; the third case combines W3 from E1 and W4 from E2. In the two 
remaining cases we combine the overlapping regions in W4 from E1 and PC 
from E2, then PC from E1 and W2 from E2.  All of the chosen cases depict 
sufficiently bright parts of the nebula and an adequate number of 
background/foreground stars which may be used for a fine co-alignment of 
the E1 and E2 images. Choosing the appropriate regions, one also has to 
find a compromise between the two conflicting factors: opting for remote 
regions of the nebula, one may minimize the bias introduced by infrequent 
plane-of-the-sky projections; on the other hand, the velocity fields of 
the remote regions may be distorted by the bow-shock interface surrounding 
the nebula (van der Sluys \& Lamers 2003).

All the available images were cleaned of cosmic rays, corrected for camera 
distortions, then iteratively shifted and rotated using a combination of 
the cross-correlation technique (e.g., {\it crossdriz}) and the 
reference-star approach (e.g., {\it imcentroid}).  In two cases we 
magnified the Planetary Camera images by adjusting their pixel scales 
which were derived by measuring distances between multiple reference 
stars. Finally, the relative shifts between out-moving parts of the nebula 
were estimated by choosing small ($\sim 100\times 100$ pixels) sections of 
the overlapping E1 and E2 images with the most obvious relative shifts and 
allowing the E1 sub-raster to move in {\it x,y} directions with 0.01-pixel 
increments. Such gradual, small-scale shifts, at some point, should 
minimize the difference (assessed via standard deviation; see Fig. 1 for 
examples of the differences) between the E1 and E2 sub-rasters, thus 
providing an estimate of the radial expansion rate.

\section {RESULTS AND CONCLUSIONS}

Long-slit spectroscopy of M1--67 established that the bulk of the nebula 
expands at $v_{\rm exp}$=42-46 km s$^{-1}$ (Solf \& Carsenty 1982; 
Sirianni et al. 1998).  In our calculations we adopt the latter value 
($v_{\rm exp}$= 46 km s$^{-1}$) which is based on a comprehensive 
modelling of the nebular expansion.  All 5 measurements (Fig 2)  point to 
a fairly uniform expansion between E1 and E2 ($\Delta t = 11.26$ yr) of 
$\Delta r=0.326 \pm 0.059$ pixels, or $\Delta r=0.^"0326 \pm 0.^"0059$.  
Assuming these expansions to be essentially radial, this translates into a 
distance to M1--67 of $d = 3.35\pm 0.67$ kpc, in perfect (although 
fortuitous) accord with the estimate from van der Hucht (2001), under the 
assumption of $M_{v}$ = --5.5 mag for the central star. The 20\% distance 
error comes from the combination of the $\sim 10\%$ 
uncertainty in the expansion velocity of the nebula and the 
above-mentioned 18\% error of the geometric expansion. The latter corresponds 
to the standard deviation of the sample of 5 independent measurements. We 
prefer this straightforward and, potentially, less biased estimate over 
the formal error analysis of the $x,y$ components of the expansion 
patterns in the individual regions. The $x,y$ errors would depend on, most 
of all, the detailed morphology of a given region, as well as different 
S/N ratios of the region's fine structures, which may lead to 
unaccountable biases. Though we selected  the relatively remote regions of M1--67 (Fig 2), there is always a chance that {\it all} of them have been affected by the plane-of-the-sky projection effect.  One may not completely disregard this possibility. However, considering the  tight clustering 
of the derived expansions (0.368,0.367,0.334,0.338,0.224 pixels), we conclude that 4 out of 5 randomly selected regions may have orientations fairly close to the plane-of-the-sky.

Such a distance unequivocally assigns a runaway status to WR124. Indeed,  the  star is moving fast, judging by the high peculiar RV$\sim 156$ km$^{-1}$ of the star and its surrounding nebula (Moffat et al 1982).  Given $b$ = +03.31$^{\circ}$, 
$d$ = 3.35 kpc leads to an elevation of $z=193\pm 39$ pc above the Galactic 
plane, or  4  times the average $|z|=49$ pc for Galactic Population I WR stars. 
Nothing significant can be said about the tangential (proper) motion of 
WR124,  since the Hipparcos data are insufficiently accurate for such 
distant  stars (Moffat et al. 1998). The $d$ = 3.35 kpc distance implies a 
physical radius  of 0.9 pc for M1--67 based on its 55$''$ radius (Solf \& 
Carsenty 1982;  Grosdidier et al. 1998), and so a dynamical age of 21,000 
yr for our  adopted 46 km $^{-1}$ expansion rate.  

On the basis of the geometric distance to WR 124/M1--67, we have 
re-derived the physical properties of WR 124 using ultraviolet, optical 
and near-infrared spectroscopic datasets from C99 
plus the model atmosphere code {\sc CMFGEN} (Hillier \& Miller 1998). Our 
analysis mimics C99, except that line blanketing by additional elements is 
incorporated, namely oxygen, neon, sulfur, argon, calcium and nickel, such 
that, in total, a factor of three more spectral lines are considered. A 
slightly higher stellar temperature of 36 kK  is obtained, while the 
reduced distance with respect to that adopted by C99 (5 kpc) yields a 
lower luminosity and mass-loss rate for  identical adopted clumping 
factors of $f$ = 0.1, with an  atmospheric hydrogen content ($\sim$15\% by 
mass) unchanged. We compare the newly derived properties of WR124 with C99 
and Hamann et al. (2006, H06) in Table~\ref{params}. Stellar masses are
estimated in each case using the mass-luminosity relation for 
hydrogen-free WR stars of Schaerer \& Maeder (1992). We have compared our 
results  with contemporary evolutionary models for single stars at solar 
composition (Meynet \& Maeder 2003). The closest agreement is found 
for an initially rotating 25 $M_{\odot}$ model after 8.6 Myr, although
this remains somewhat more luminous and hydrogen-rich than WR 124, with 
$\log L/L_{\odot}$=5.47 and H/He=1.5 ($X_{\rm H}$ = 27\%).

\begin{center}
\begin{deluxetable}{llll}
\tablecaption{Comparison between physical and wind properties of WR124 
(WN8) derived by Crowther et al. (1999, C99), Hamann et al. (2006, H06) 
and the present study.\label{params}}
\tablewidth{0pc}
\tablecolumns{3}
\tablehead{\colhead{Study} & \colhead{C99} & \colhead{H06} & \colhead{This 
study} }
\startdata
$T_{\ast}$ (kK) & 32.7 & 44.7 &  35.8 $\pm$2 \\
$R_{\ast}$ (R$_{\odot}$) & 18.0  & 16.7 & 10.1 $^{+2.8}_{-2.6}$  \\
%
%
$\log L/L_{\odot}$ &   5.53  &  6.0 & 5.18 $^{+0.2}_{-0.24}$  \\
$M/M_{\odot}$         &  14:  &  33: & 9 $\pm$ 2.5 \\
$v_{\infty}$ (km\,s$^{-1}$) & 710 & 710 & 710 \\
$f$            & 0.1 & 0.25 & 0.1 \\
$\dot{M}$ ($M_{\odot}$\,yr$^{-1}$) & 10$^{-4.7}$& 10$^{-4.1}$ &
10$^{-4.95 \pm 0.15}$  \\
H/He ($X_{\rm H}$) & 0.7 (15\%) & 0.6 (13\%) & 0.75 $\pm$ 0.1 (15$\pm$2\%) 
\\
\\
$m_{\rm v}$ (mag) & 11.58   & 11.58 & 11.58   \\
$A_{\rm v}$ (mag)    & 4.4:  & 4.1  &  4.3 $\pm$ 0.1   \\
$d$ (kpc)         & 5    & 8.4 & 3.35 $\pm$ 0.67    \\
$M_{\rm v}$ (mag) & --6.3  & --7.2  & --5.3 $^{+0.4}_{-0.5}$    \\
\enddata
\tablecomments{Stellar mass estimates are obtained from the
mass-luminosity relation for hydrogen-free WR stars (Schaerer \& Maeder 
1992).}
\end{deluxetable}
\end{center}

Our target is unlikely to undergo a GRB since such events are 
predominantly found in faint, metal-poor host galaxies (e.g. Modjaz et al.
2008). Nevertheless, the proven runaway status  of a Population I WR star 
may provide some insights for  interpretation of  the afterglows coming 
from a specific category of GRBs.  Let us assume that  a rapidly moving 
massive star produces a long GRB. The 
circumburst media  are usually probed by modeling the temporal decay of 
afterglows (Chevalier 
\& Li 2000), with preference given to the late (0.1 day and more) phases 
of an afterglow, in order to lessen the impact of any prompt-emission 
components. As a consequence, typical time-spans used in the recent models 
of afterglows cluster around $10^5$ sec (e.g., Starling et al 2008).  
Hence, if an afterglow points to a seemingly homogeneous environment in 
the {\it immediate} surroundings of the burst, the size of the wind-filled 
cavity should not exceed a few light-days. Otherwise, in an unbound, 
freely expanding stellar wind the density profile would fall off 
proportionally to $r^{-2}$. In the case of WR 124 the density of the 
ejected M1--67 nebula falls off as $r^{-0.8}$ (Grosdidier et al. 1998). 
Considering the high RV of the ejecta, one may assume that the nebula was 
created when the star already gained high spatial velocity. Hence, the 
nebula, most likely, travels along with the central star. However, the 
star's immediate surroundings (out to at least light-weeks: the star has 
been known as a WR for more than 50 years) are filled by a fast stellar 
wind. This would result in a clear 'wind-imprinted' afterglow after a 
hypothetical GRB. Therefore where do the $\sim$constant-density profiles 
come from?

A massive runaway star with a strong wind, moving through the ISM at 
supersonic velocity creates a leading bow shock (e.g., Vela X-1: 
Kaper et al. 1997). Beyond the wind-filled cavity and the bordering 
bow-shock, both occurring relatively close to the central star, one may 
expect to see an unperturbed, 'homogeneous' interstellar environment. The 
bow-shock stand-off distance, $d_{s}$, can be expressed as (Huthoff \& 
Kaper 2002):
\[d_s=\frac {0.9 pc}{(v_\star /50{\rm km\,s}^{-1})}\sqrt {\frac { 
(\dot {M}/10^{-6}M_\odot )\times (v_{\infty }/10^{3}
km\,s^{-1})}{(n /{\rm cm}^{-3})}}\]
where $\dot {M}$ is the mass-loss rate and $v_{\infty}$ is the terminal wind velocity.

The peculiar spatial velocity of WR 124 may reach  $v_\star \sim 200$ km 
s$^{-1}$ (e.g. Moffat et al 1982).  However, the typical [high] mass-loss rate of a 
Galactic WN8 star prevents $d_s$ from dropping much below a few 
light-months, instead of the required light-days. Indeed, van der Slyus 
and Lamers (2003) find that M1--67 is confined by a bow shock with 
[model-dependent] $d_s=1.3$ pc. The situation changes dramatically once we 
invoke the fact that the long GRBs preferentially occur in low-metallicity 
environments. In this case one must take into account the dependence of 
the mass-loss rate and the terminal velocity of the wind on the 
metallicity of the host galaxy: $\dot M\sim Z^{0.83}$ (Mokiem et al. 
2007), and $v_\infty \sim Z^{0.13}$ (Leitherer et al. 1992). In addition, 
the rest-frame spatial interstellar densities, $n$, of the regions hosting GRBs 
usually exceed the corresponding densities in the disk of the Milky Way 
Galaxy by 1-2 orders of magnitude (Nardini et al. 2010). A typical WN8 
wind with $\dot M\sim 5\times 10^{-6}M_\odot {\rm yr}^{-1}$ and 
$v_\infty\sim10^3$ km s$^{-1}$ (Hamann et al. 2006), once appropriately scaled 
down for a $Z=10^{-3}Z_\odot$ environment to $\dot M\sim 1.6\times 10^{-8}M_\odot {\rm yr}^{-1}$ 
and $v_\infty \sim 400$ km s$^{-1}$, would create 
a rather small wind-blown cavity around a fast-moving star with $d_s$ of 
the order of a few light-days. This should increase the chance of at least 
one of the ensuing GRB jets to promptly leave the wind-filled cavity and 
induce an afterglow which could be perceived as coming from an  approximately
constant-density environment unperturbed by a stellar wind.

WR 124 itself is not likely to end its life as a GRB. However, it may explode as a core-collapse SN, either a Type IIb 
or Ib. Both flavours of core-collapse SNe are  closely 
associated with  star-forming  regions and agglomerates of massive stars 
(Smartt 2009).  The possible association of a dense ejection nebula, such
as M1--67, with the immediate supernova progenitor also has relevance to 
the  interpretation of the interaction between the SN blastwave and the 
inner  circumstellar material (Chevalier 1982), as, for example, measured 
by radio observations (Weiler et  al. 2002). 

Pondering the origin of the high spatial velocity of WR 124, we note that 
the star moves away almost radially (van der Sluys \& Lamers 2003). The absence of 
any rich cluster/association in the star's vicinity, as well as the 
relatively low initial stellar mass (the newly derived luminosity suggests 
the current mass of $\sim 9\,M_\odot$) point to a modest parent 
cluster, if any. However, the high RV of the star 
calls for  strong dynamical interactions in a dense core of a rich stellar cluster, contrary to 
observations. Hence, it seems more likely
that WR 124 has gained the runaway status following the recoil after a SN 
explosion in a close binary system.

We conclude that the relatively assumption-free, geometric distance to WR 
124 re-inforces the runaway status of the star. The implied 
high peculiar velocity should result in a large displacement of WR 
124 from its place of origin, thus potentially moving the runaway into a 
region with relatively unperturbed ISM. The latter may help to 
explain the occurrence of a substantial fraction of GRB afterglows in 
regions with a seemingly constant-density circumburst medium, contrary to 
the expectations of a $\sim r^{-2}$ density fall-off for a typical 
massive-star wind. However, an appropriately small ($\sim$light-day) 
wind-filled cavity may only be formed by a massive runaway star born in a 
low-metallicity environment, $Z\leq 10^{-3}Z_\odot$: cf. $Z=10^{-2}Z_\odot$ 
for the host of GRB 050730 (Chen et al. 2005).  Hence, this particular category of GRB 
may emerge exclusively from regions 
with a {\it very} low metallicity content, thus  conforming  to the general 
conclusion of Han et al. (2010) that long GRBs should occur in a 
relatively pristine ISM.

\acknowledgments This study was based on observations made with the 
NASA/ESA Hubble Space Telescope and obtained from STScI, 
which is operated by AURA Inc., under NASA contract NAS 5-26555. Support 
for Program number GO11137 was provided by NASA through a grant from the 
STScI, which is operated by AURA Inc., under NASA contract 
NAS 5-26555. AFJM is grateful for financial assistance to NSERC (Canada) 
and FQRNT (Qu\'ebec).

\begin{figure} \plotone{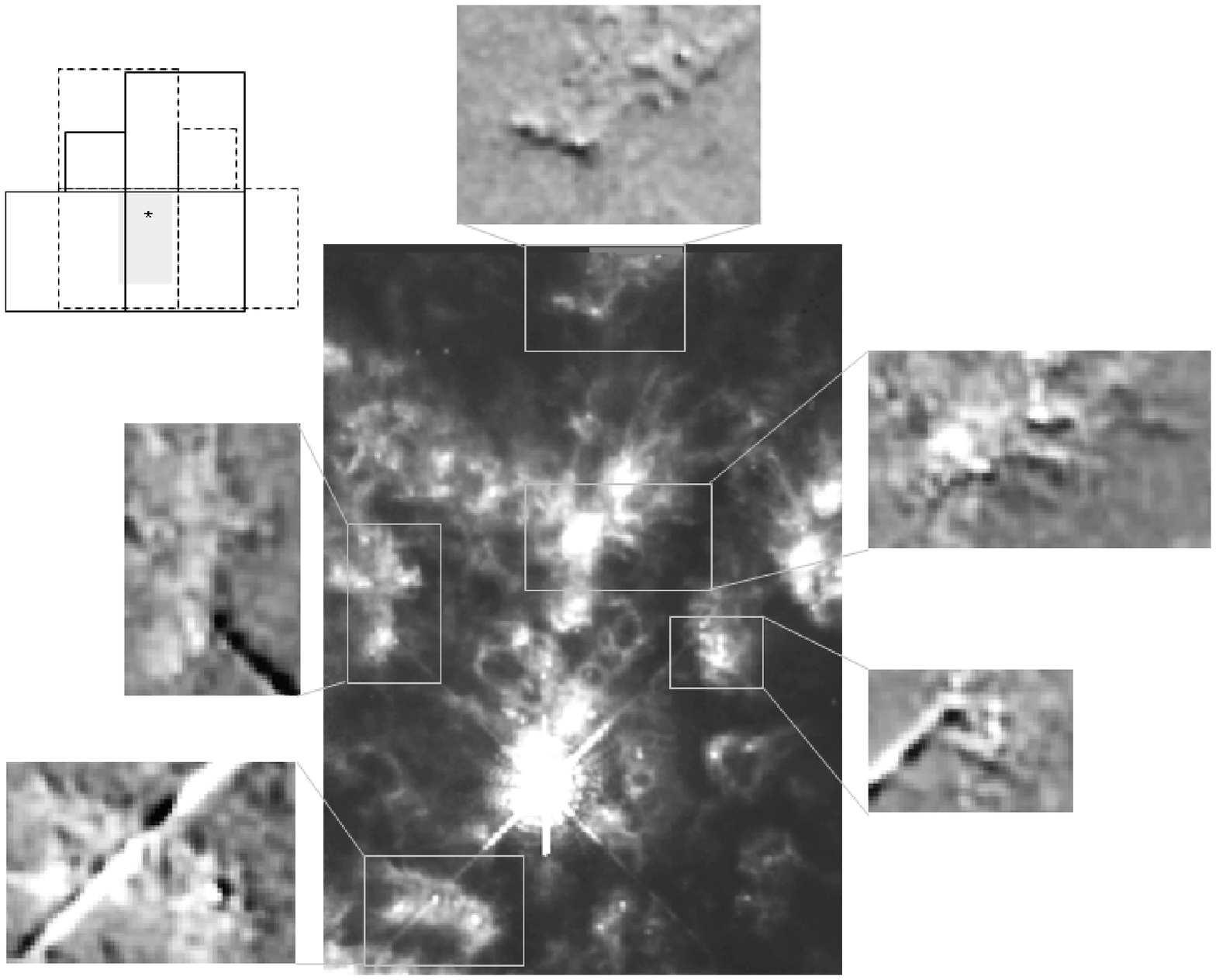} \caption{The sketch in the upper-left 
corner shows the orientation of the E1 (full lines separating the W2,W3,W4 
and PC devices) and E2 (dashed lines) HST images taken with the 656N filter. The central star is placed 
in W3 for both epochs. The shaded area in the sketch corresponds to the 
part of the W3 (E2) image shown in the main section of the figure. The 
magnified parts of the W3 image show the E2-E1 differences, where the 
proper motions are most easily detected via relatively strong, sharp 
nebular edges.\label{fig1}}\end{figure}

\begin{figure} \plotone{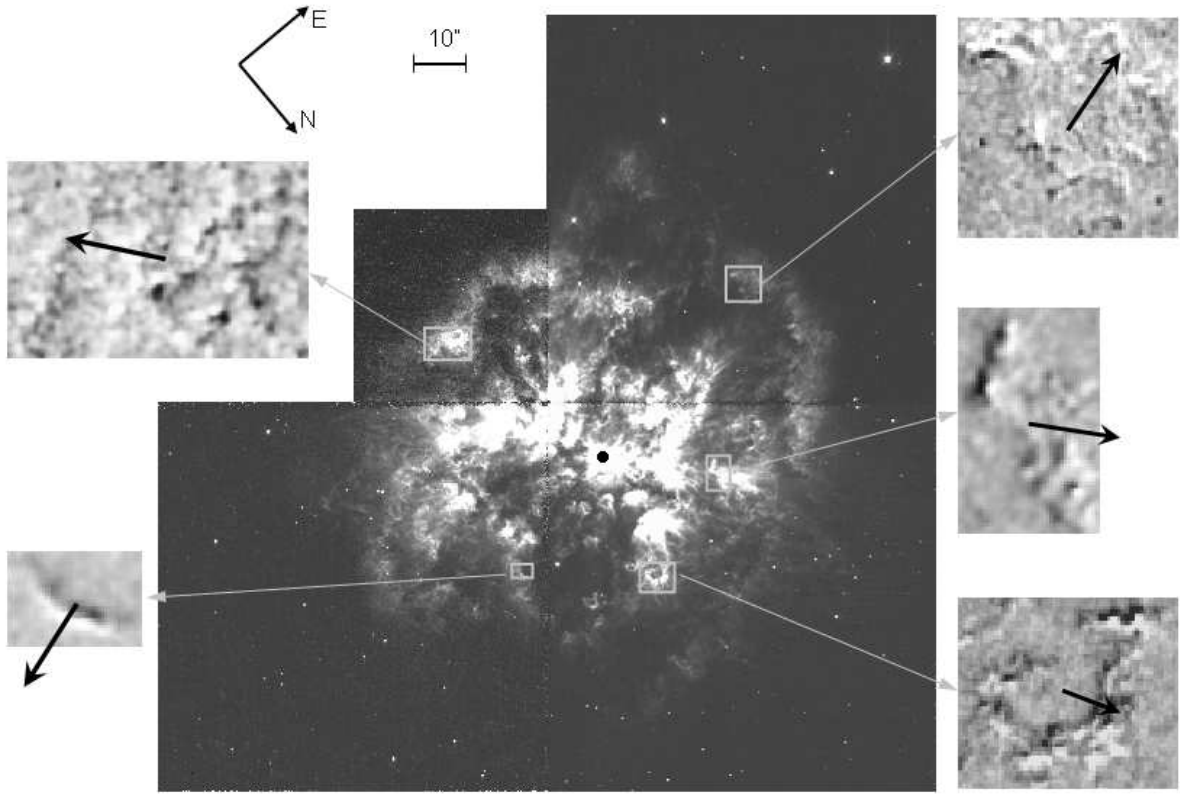} \caption{The E2 656N image of M1--67 with the 
central star marked by a dot.  The enlarged areas show the E2-E1 
difference images used to estimate the geometric expansion of the nebula. The 
direction and size of the black arrows depict the measured proper motions 
of the corresponding parts of the nebula. The scale and 
orientation of the image are provided in the upper left corner.\label{fig2}}\end{figure}  
\clearpage

\end{document}